\documentclass[aps,prl,superscriptaddress,twocolumn,10pt,nobibnotes]{revtex4-1}
\usepackage{graphicx,float}
\usepackage{bm}
\usepackage[colorlinks=true]{hyperref}
\usepackage[dvipsnames]{xcolor}
\usepackage{amsmath,amssymb} 
\newcommand*{\vk}{\mathbf{k}}

\begin{document}

\title{\mbox{Topological magnon-polaron transport in a bilayer van der Waals magnet}}
\author{Zhi-Xing Lin}
\affiliation{Department of Physics, Princeton University, Princeton, New Jersey 08544, USA}
\author{Shu Zhang}
\email{suzyccheung@gmail.com}
\affiliation{Max Planck Institute for Physics of Complex Systems, 01187 Dresden, Germany}

\date{\today}

\begin{abstract}
The stacking of intrinsically magnetic van der Waals materials provides a fertile platform to explore tunable transport effects of magnons, presenting significant prospects for spintronic applications. The possibility of having topologically nontrivial magnons in these systems can further expand the scope of exploration. In this work, we consider a bilayer system with intralayer ferromagnetic exchange and a weak interlayer antiferromagnetic exchange, and study the topological magnon-polaron excitations induced by magnetoelastic couplings. Under an applied magnetic field, the system features a metamagnetic transition, where the magnetic ground state changes from antiparallel layers to parallel. We show that the metamagnetic transition is accompanied by a transition of the topological structure of the magnon polarons, which results in discernible changes in the topology induced transport effects. The magnetic-field dependence of the thermal Hall conductivity and spin Nernst coefficient is analyzed with linear response theories.
\end{abstract}

\maketitle

Intrinsic magnetism in van der Waals (vdW) materials has opened new avenues to explore magnetic phenomena in low dimensions~\cite{burch2018,jiang2021review,zhang2021two,wang2022magnetic,kurebayashi2022magnetism,liu2023vdWM,tang2023reivew}. Advances in this area have led to a variety of platforms for studying quantum and topological aspects of magnetism. As magnon excitations in vdW magnets are gathering increasing interest, one particular focus is their potential for exhibiting nontrivial topology~\cite{owerre2016,zhu2021topological,zhuo2023review}, with a few prominent material examples found in the transition metal trihalide family~\cite{chen2018cri3,chen2021cri3,cai2021CrBr3,Nikitin2022crbr3,zhang2021vi3}.
Topological magnons can lead to spin and heat transport in the transverse channel~\cite{onose2010,matsumoto2011prl,matsumoto2011,mook2014,hirschberger2015,zyuzin2016,Kawano2019,zhang2019,park2020,to2023giant}, which can be harnessed for spin current control and thermal management~\cite{wang2018review,ghader2022theoretical}. Their magnetic nature and bosonic statistics offer high tunability with magnetic field and temperature.~\cite{mook2014,go2019,szhang2020,Shen2020,neumann2022,ma2022,hu2022,Soenen2023,klogetvedt2023tunable}.

In addition, magnon-phonon couplings, which commonly exist and can be significant in vdW magnets~\cite{du2019FGT,Bazazzadeh2021Magnetoelastic,Liu2021FPSe,mai2021MPSe,kozlenko2021crbr3,esteras2022crsbr,hovancik2022vi3,delugas2023cri3,luo2023FPSe,lyons2023crcl3}, have been proposed as a route to induce nontrivial topology in collinear magnets without relying on anisotropic magnetic interactions~\cite{go2019,szhang2020}. 
The corresponding topological excitations are magnon polarons with a hybridized magnon and phonon character.
Recent experimental results on thermal Hall effects in VI$_3$~\cite{zhang2021vi3}, {FePS$_3$\cite{Vaclavkova2021magnon}, FeCl$_2$~\cite{xu2023}, and {Fe$_2$Mo$_3$O$_8$ \cite{bao2023direct}, for example, have shown strong evidence of this mechanism. 

Parallelly, the versatile stacking of vdW magnetic layers largely enriches their thermodynamic and transport behaviors~\cite{gong2019heterostructure,choi2022heterostructure}. 
One interesting and simple scenario is the stacking of antiferromagnetically coupled ferromagnetic layers, which is similar to synthetic antiferromagnets~\cite{duine2018synthetic},
whereas smaller and lighter devices are achievable with atomically thin vdW magnets.
The magnetic properties of these systems can show a strong layer dependence~\cite{yang2021mbt,ovchinnikov2021intertwined}. Also, in contrast to the spin-flop transition in conventional antiferromagnets, a metamagnetic transition can be induced with a relatively low magnetic field due to the weak interlayer coupling, where entire layers antiparallel to the field are flipped, causing a drastic change in the total magnetization. In the case of FeCl$_2$~\cite{kenan1969meta,jacobs1967meta}. this transition is accompanied by an abrupt sign change in the thermal Hall conductance~\cite{xu2023}. 

\begin{figure}[b]
\vspace{-10pt}
    \centering
    \includegraphics[width=0.5\textwidth]{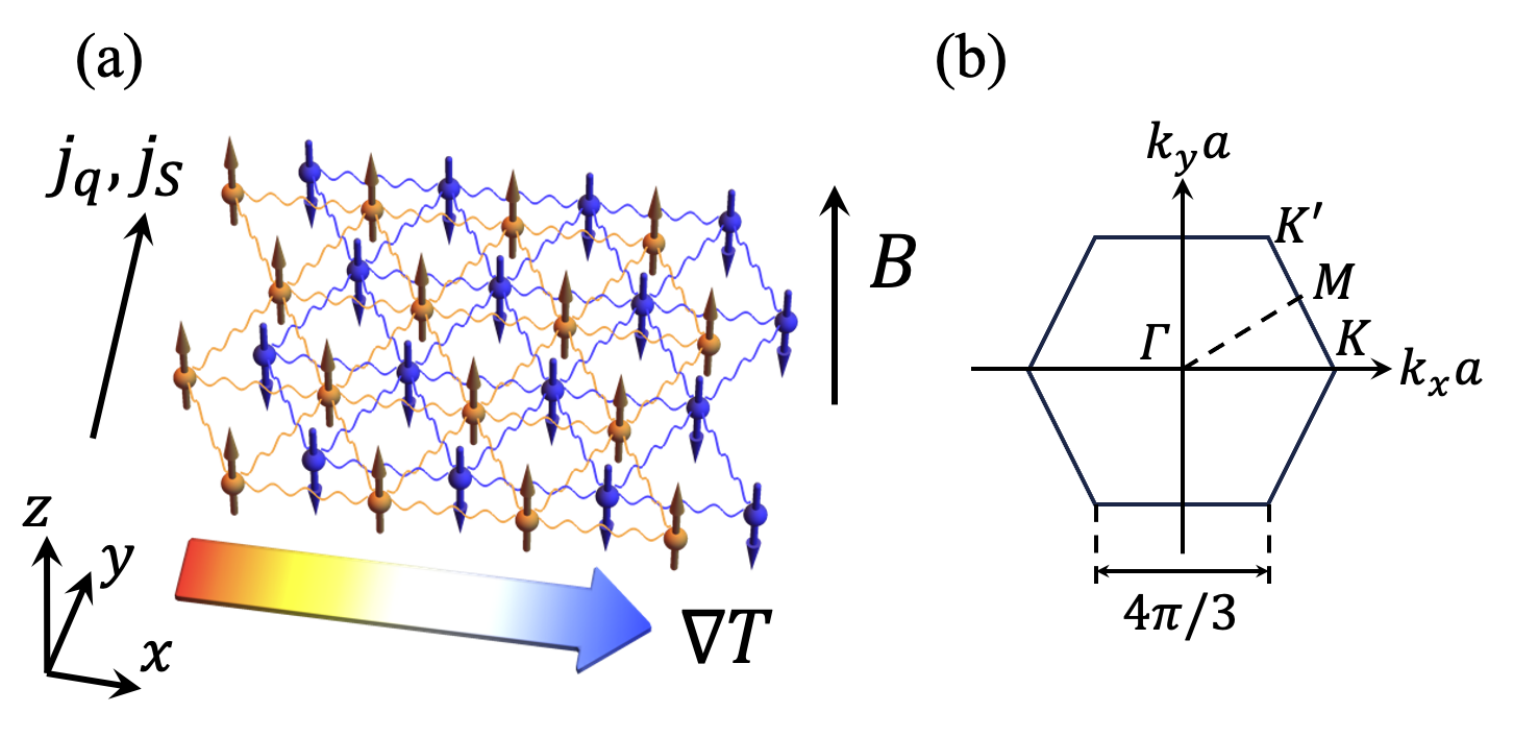}
    \caption{(a) 
    The triangular-lattice bilayer studied in this work. Topological magnon polarons in the system can give rise to thermal Hall and spin Nernst effects.
    The system features a metamagnetic transition in an external field $B$, switching from an antiparallel layered ground state to a parallel one.
    (b) The first Brillouin zone, where $a$ is the lattice constant.
    }
    \label{fig:lattice}
\end{figure}

To reveal the topological behavior of magnon polarons and the associated transport properties of such a system, we theoretically study a bilayer model, where two ferromagnetic triangular-lattice layers are coupled with a weak antiferromagnetic interaction [Fig.~\ref{fig:lattice}~(a)].
We find that the metamagnetic phase transition triggers a topological transition of the magnon-polaron bands. This leads to a sign flip in thermal Hall conductance, but not in the spin Nernst coefficient, which instead exhibits a smooth sign change in the antiferromagnetic phase. The topological transition in the magnon-polaron band structure across the critical magnetic field (Fig.~\ref{fig:topology}) and the qualitative field dependence of the transverse thermal and spin transport (Fig.~\ref{fig:transport}) constitute the main results of this work.
Our discussions can be generally applicable in vdW magnets with similar magnetic structures and magnon-phonon couplings. 

We demonstrate our ideas with the example of a triangular-lattice bilayer. As shown in Fig.~\ref{fig:lattice}~\textcolor{red}{(a)}, the stacking geometry has each atom located at the center of a triangle in the other layer, as is typical in transition metal dihalides. The following model Hamiltonian is considered: 
\begin{equation}
    H = H_\text{m} + H_\text{e} + H_\text{me},
\label{eq:full-Hamiltonian}
\end{equation} 
which entails the magnetic, elastic, and magnetoelastic terms.

\begin{figure*}[t]
  \centering
    \centering
    \includegraphics[width = \linewidth]{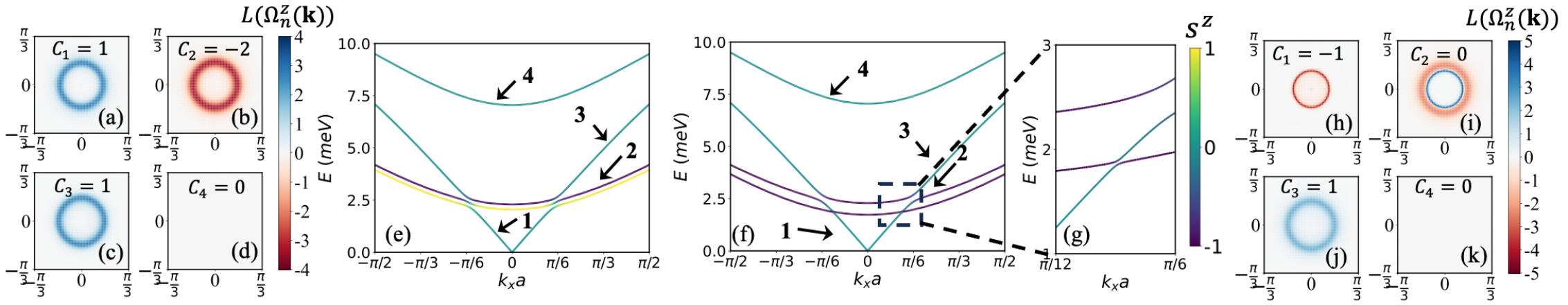} 
    \caption{Magnon-polaron bands and their topological structure. Berry curvatures (a)-(d) for the of the hybridized magnon-phonon bands (e) in the antiparallel phase at field $h=0.12\ \mathrm{meV}$, and in parallel phase (f)-(k) at $h=0.36 \ \mathrm{meV}$, where a small staggered field $\Delta h$ is turned on to gap the lower crossing point. The bands are colored by the spin weight they carry. Berry curvatures are plotted in the log scale $L(\Omega_n^z(\vk))=\mathrm{sign}(\Omega_n^z)\log (1+|\Omega_n^z|)$. See the main text for the parameters used.
    }
    \label{fig:topology}
    \vspace{-10pt}
\end{figure*}

The magnetic interactions involve an intralayer nearest-neighbor ferromagnetic exchange $J$ and an interlayer nearest-neighbor antiferromagnetic exchange $J'$, where $0 < J' \ll J$ due to the vdW nature of the material. 
In addition, the magnetic Hamiltonian includes a 
an easy-axis anisotropy $K>0$, and the Zeeman coupling to an external magnetic field perpendicular to the layer plane $\mathbf{B} \!=\! B \hat{\mathbf{z}}$. 
In the absence of the external field, the Hamiltonian favors a N{\'e}el magnetic order along the $\hat{\mathbf{z}}$ axis, consisting of two antiparallel layers, whereas the spins within each layer are ferromagnetically aligned. As the field along $\hat{\mathbf{z}}$ increases, the system features a metamagnetic spin-flip transition, where both layers become aligned with the field. Since the antiferromagnetic interlayer coupling is weak, the spin-flip transition happens at a relatively low critical field,  $h_c {=\mu_B g_\parallel B_c} \sim 3 S J'$ at the zero temperature mean-field level, before the typical spin-flop transition can happen.


At temperatures much lower than the N{\'e}el temperature, the magnetic excitations can be described by magnons via a Holstein-Primakoff transformation~\cite{Holstein1940} with respect to the ground state spin configurations. Under small magnetic fields $h<h_c$, the antiparallel ground state has 
$\mathbf{S}_{i\in A}=-\mathbf{S}_{i\in B}=S\hat{\mathbf{z}}$, where $A$ and $B$ denote the upper and lower layer, respectively. 
We obtain the magnon Hamiltonian in the momentum space:
\begin{equation}\label{eq:ap-magnon}
    H_\text{m}^\text{AP} = \sum_\vk \epsilon^m_+ a^\dagger_\vk a_\vk + \epsilon^m_- b^\dagger_\vk b_\vk + S J' \gamma_\vk \left( a^\dagger_\vk b^\dagger_{-\vk}+a_{-\vk}b_\vk \right),
\end{equation}
where the detailed expressions of $\epsilon^m_\pm, C_\vk, \gamma_\vk, a_\vk, b_\vk$ are given in the Supplementary Material.

Above the metagmagnetic transition field $h>h_c$, the bilayer ground state is parallel to $\mathbf{S}_{i\in A} = \mathbf{S}_{i\in B} = S\hat{\mathbf{z}}$.
The magnon Hamiltonian is instead
\begin{equation}
    H_\text{m}^\text{P} = \sum_\vk \epsilon^m_+ \left( a^\dagger_\vk a_\vk + b^\dagger_\vk b_\vk \right) + J'S \left(\gamma_\vk a^\dagger_\vk b_{\vk}+\gamma_\vk^\ast a_{\vk}b^\dagger_\vk \right),
\end{equation}

For the elastic part, we only consider the out-of-plane displacement $u_{i,\vk}^z$ , as it couples to the magnon operators in the lowest order in the magnon-phonon coupling, as shown later,
\begin{equation} 
\begin{aligned}
    H_\text{e} =& \sum_{\vk} \sum_{i=A, B}\left[\frac{1}{2M} p^z_{i,\vk}p^z_{i,-\vk}+\frac{1}{2} \left( \lambda C_\vk+3\lambda' \right) u^z_{i,\vk}u^z_{i,-\vk} \right] \\
    -& \sum_{\vk}\frac{1}{2} \lambda' \left(\gamma_\vk u^z_{A,-\vk} u^z_{B,\vk} + \gamma_\vk^\ast u^z_{A,\vk} u^z_{B,-\vk} \right),
\end{aligned}   
\end{equation}
{where $M$ is the atom mass,  $p^z_{i,\vk} \!=\! M \dot{u}^z_{i,\vk}$ is the kinetic momentum, $\lambda$ is the bond stiffness for intralayer nearest neighbors and $\lambda'$ for interlayer.}

The phonons in the bilayer split into an acoustic branch and an optical branch due to the interlayer bond stiffness, which can still be sizable because the Coulombic interactions between atoms are long-range~\footnote{The interlayer stiffness associated with \textit{in-plane} displacements (e.g. sliding motion) can of course be much smaller}, in contrast to the fast decreasing magnetic exchange with distance. The optical branch is thus gapped away from the magnons, while the gapless acoustic branch can cross through the magnon bands with a typically steeper dispersion.

The magnon-phonon interactions introduce a topologically nontrivial hybridization at the crossing.
We discretize the symmetry-allowed magnetoelastic couplings~\cite{kittel1958} on the triangular lattice, and consider the terms linear in magnon operators~\cite{thingstad2019,go2019,szhang2020}.
In the antiparallel phase, this leads to the magnon-phonon coupling
\begin{equation}
\begin{aligned}
H_\text{me}^\text{AP} = \kappa \sqrt{2S^{3/2}} \sum_\vk &\left[(-\nu_{1\vk}+\nu_{2\vk}) \, a^\dagger_\vk u_{A,\vk} \right.\\
 & \left.+ \, (\nu_{1\vk}+\nu_{2\vk}) \, b^\dagger_\vk u_{B,\vk} \right] + \text{h.c.}, 
\end{aligned}
\end{equation}
{where $\kappa$ is the magnetoelastic coupling constant and $\nu_{1,2\vk}$ are provided in the Supplementary Material.

For the parallel bilayer, on the other hand, we have
\begin{equation}
\begin{aligned}
H_\text{me}^\text{P} & =\kappa \sqrt{2S^{3/2}} \sum_\vk (-\nu_{1\vk}+\nu_{2\vk})(a^\dagger_\vk u_{A,\vk} \!+\! b^\dagger_\vk u_{B,\vk}) \!+\! \text{h.c.}.
\end{aligned}
\end{equation} 
The effect of these couplings is the most prominent as a first-order perturbation at the degenerate point of a magnon and a phonon, where they lift the band degeneracy and open a gap. 
We neglect the interlayer magnetoelastic terms, which are much weaker and do not affect our results qualitatively.

We proceed to analyze the band topology of the magnon-polaron excitations in the antiparallel and parallel phases.
To treat the boson pairing terms such as those in Eq.~(\ref{eq:ap-magnon}), we reform the total Hamiltonian~(\ref{eq:full-Hamiltonian}) at the quadratic order in the Bogoliubov-de Gennes (BdG) form
$H \!=\! (1/2) \sum_\vk \phi_\vk^\dagger \mathcal{H}_\vk \phi_\vk$ with the basis 
$\phi_\vk \!=\! (a_\vk, b_\vk, c_{A,\vk}, c_{B,\vk}, a^\dagger_{-\vk}, b^\dagger_{-\vk}, c_{A,-\vk}^\dagger, c_{B,-\vk}^\dagger)^T$.  Here, we have taken the transformation
$u_{i,\vk}\!=\!\sqrt{\hbar/2}/(\lambda M)^{-1/4}(c_{i,\vk} + c_{i,-\vk}^{\dagger})$ and 
$p_{i,\vk} \!=\!-i (\lambda M)^{1/4} \sqrt{\hbar/2} \, (c_{i,-\vk} - c_{i,\vk}^{\dagger})$, $i=A, B$, such that $c_{i, \vk}$ and $c_{i, \vk}^\dagger$ operators obey the bosonic commutation relation, which is then preserved in the diagonalization using Colpa's method~\cite{COLPA1978,Maestro2004,park2019,Li2020,Bazazzadeh2021,Bazazzadeh2021Magnetoelastic}.
The resultant eigenenergies $E_n(\vk)$ and corresponding eigenvectors $|n,\vk \rangle$ are particle-hole symmetric and satisfy the normalization condition with respect to matrix $g=\text{diag} \{+1,+1,+1,+1,-1,-1,-1,-1\}$,
\begin{equation}
    \langle m,\vk|\mathcal{H}_\vk|n, \vk\rangle = E_n(\vk) \delta_{nm}\; \text{and} \; \langle m,\vk|g|n, \vk\rangle = g_{nm}.
\end{equation}
As shown in Fig.~\ref{fig:topology}~(e, f), in the antiparallel bilayer, the magnetoelastic interaction hybridizes the two magnons and the acoustic phonon and induces avoided crossings. In the parallel bilayer, however, the degeneracy at the crossing point remains between the lower magnon branch and the acoustic phonon~\footnote{This can be shown by diagonalizing the magnon and phonon Hamiltonians separately, and observe that the magnetoelastic couplings do not couple them.}.
For the band topology to be well defined, we introduce a small staggered magnetic field $\Delta h$, which results in a difference in the magnon chemical potential between the two layers. This breaks the inversion symmetry and opens a gap. The limit of $\Delta h \to 0$ is taken in the later study of transport effects.

The topological nature of the bands is characterized by the generalized Berry curvature  of the BdG Hamiltonian~\cite{Li2020,park2019},
\begin{equation}
    \Omega_n^z (\vk) = -\sum_{m\ne n} \frac{2\text{Im} \langle n| v_x |m\rangle \langle m|v_y|n \rangle g_{nn} g_{mm}}{[g_{nn} E_n(\vk)-g_{mm} E_m(\vk)]^2} ,
\end{equation}
where $\textbf{v} = \hbar^{-1} \nabla_\vk \mathcal H$. The results are shown in Fig.~\ref{fig:topology}.
Integrating the Berry curvature over the first Brillouin zone, we find the Chern numbers of the bands from low to high energies are $(1, -2, 1, 0)$ in the antiparallel phase, and $(-1, 0, 1, 0)$ in the parallel phase. Therefore, the metamagnetic transition is accompanied by a topological transition in this bilayer system. 

Under certain approximations, the topological structure can be understood through a SU(3) formalism following Ref.~\cite{szhang2020,Go2022}. Though $s^z$ is no longer a good quantum number in the presence of magnetoelastic couplings, we can still gain some intuition by labeling it. The structure of the Chern numbers is linked to the spin label.  In the antiparallel phase~[Fig.~\ref{fig:topology} (a-e)], assuming the magnetic field is moderately large to well separate the two magnon modes, the Berry curvatures concentrate near the avoided crossings~\cite{szhang2020,Shen2020}. We may associate a $(1,-1)$ structure with band 1 and 2 for the hybridization between the acoustic phonon and the spin-up magnon, and $(-1,1)$ to band 2 and 3 for the spin-down case. Together, they result in the $(1, -2, 1)$ distribution among the three bands.  In the parallel phase with only spin-down magnons, we thus have $(-1,1)$ overlaying with $(-1,1)$, yielding $(-1, 0, 1)$.



In the rest of this work, we focus on the consequent topological transport of heat and spin by performing linear-response calculations in the BdG form.
The nontrivial topology of the hybridized magnon-phonon excitations gives rise to thermal Hall and spin Nernst effects, which refer to a transverse heat and spin current, respectively, driven by a longitudinal temperature gradient. 
The thermal Hall conductance is given by~\cite{matsumoto2011,matsumoto2011prl}, 
\begin{equation}
    \kappa_T = -\frac{k_B^2 T}{\hbar}\sum_{n=1}^4 \int d^2\vk \,\Omega_n^z(\vk) c_2[\rho_n(\vk)],
\end{equation}
where $c_2(\rho)= (1+\rho)\ln^2 [(1+\rho)/\rho]-\ln^2\rho -2\mathrm{Li}_2(-\rho)$ and $\rho_n(\vk)=1/\left[\exp(E_n(\vk)/k_BT)-1\right]$ is the Bose-Einstein distribution. Only particle bands contribute to the summation. 
The spin Nernst coefficient 
\begin{equation}
    \alpha_{xy}^S = 2 k_B \sum_{n=1}^4 \int d^2 \vk \, \Omega_n^{s,z}(\vk) c_1[\rho_n(\vk)],
\end{equation}
where $c_1(\rho)= (1+\rho)\ln (1+\rho) - \rho \ln \rho$,
can be expressed in terms of the generalized spin Berry curvature~\cite{zyuzin2016,Li2020}
\begin{equation}\label{eq:spin-bc}
    \Omega_n^{s,z} (\vk) = -\sum_{m\ne n} \frac{2\text{Im} \langle n| \theta_x |m\rangle \langle m|v_y|n \rangle g_{nn} g_{mm}}{[g_{nn} E_n(\vk)-g_{mm} E_m(\vk)]^2}.
\end{equation}
Here, we have defined the spin current operator $\bm{\theta} = \frac{1}{4} (s^z g \mathbf{v}+ \mathbf{v} g s^z)$ and the spin operator $s^z= - \sum_\vk (a^\dagger_\vk a_\vk +a_{-\vk} a^\dagger_{-\vk}) \pm (b^\dagger_\vk b_\vk +b_{-\vk} b^\dagger_{-\vk})$ ($+$ for the parallel case and $-$ for antiparallel).

For the quantitative calculations, we choose a set of parameters in accordance with the vdW magnetic material $\mathrm{FeCl}_2$, where $\mathrm{Fe}^{2+}$ ions have spin $S=1$. 
The intralayer exchange $J=0.6\ \mathrm{meV}$ and magnetic anisotropy $K=2\ \mathrm{meV}$ reproduce a magnon gap $\sim 2 \ \mathrm{meV}$ and a  magnon bandwidth $\sim 5\ \mathrm{meV}$, and the magnetoelastic coupling strength $\tilde{\kappa} = \sqrt{\hbar}  S^{3/2} \kappa (\lambda M)^{1/4} = 0.18 \ \mathrm{meV}$ yields a energy splitting $\sim 0.3$~meV at the magnon-phonon crossing, as measured in neutron scattering~\cite{birgeneau1972,Lovesey1974,BALUCANI1986}. The interlayer exchange $J'=0.09 \ \mathrm{meV}$ and g-factor $g_\parallel \sim 4$ are used, consistent with the metamagnetic critical field $\sim 10^4$Gs~\cite{carrara1969}. For the elastic sector, we set $\hbar \sqrt{\lambda / M} = 3.8$~meV for the acoustic phonon velocity to be $\sim 2.1$~km/s~\cite{YELON1974}, and the ratio of bond stiffnesses $\lambda'/\lambda=0.58$ to be inversely proportional to the geometric distance. 

\begin{figure}[b]
\vspace{-10pt}
    \centering
    \includegraphics[width=0.48\textwidth]{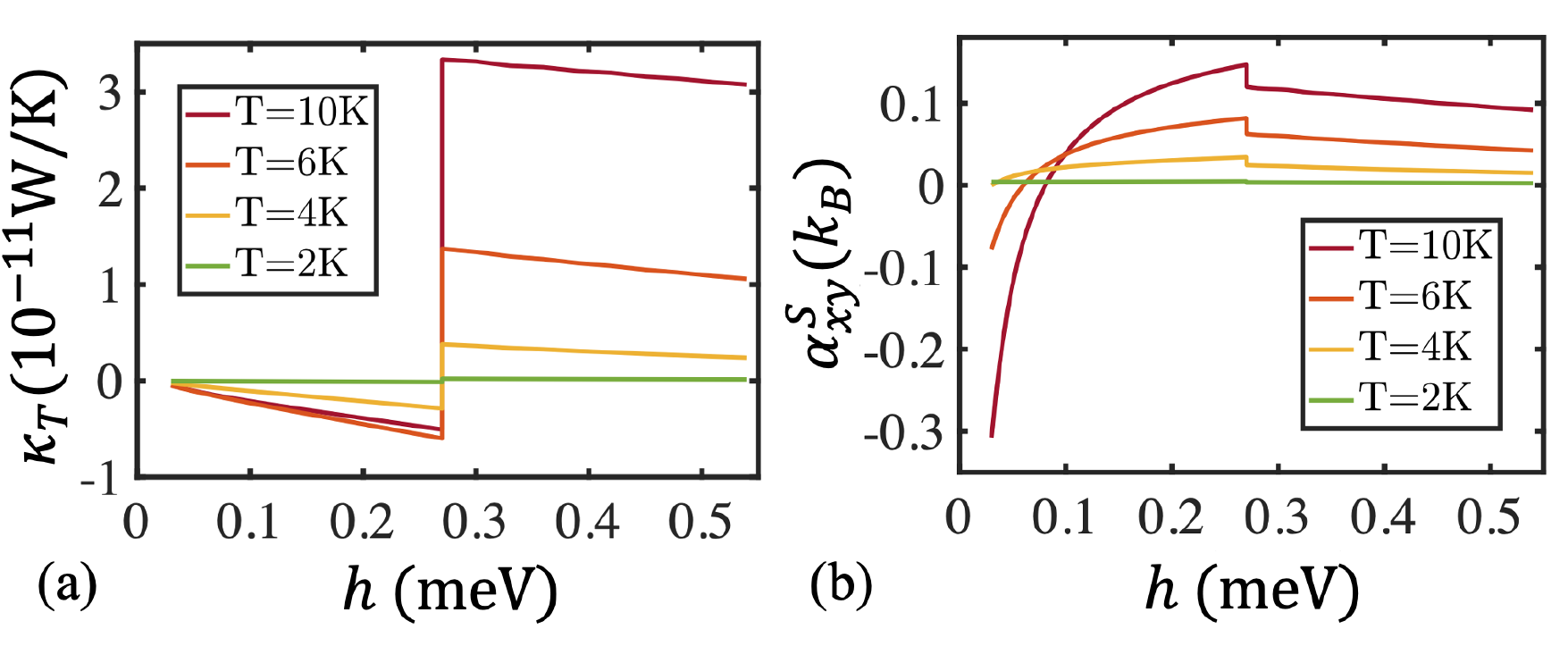}
    \caption{The external field $h$ dependence of (a) the thermal Hall conductance $\kappa_T$ and (b) the Nernst spin coefficient $\alpha_{xy}^S$ at different temperatures.}
    \label{fig:transport}
\end{figure}

Figure~\ref{fig:transport} shows the dependence of thermal Hall conductance $\kappa_T$ and the spin Nernst coefficient $\alpha_{xy}^S$ on the magnetic field at temperatures well below the N{\'e}el temperature. As the bilayer undergoes the metamagnetic transition, a sharp change in magnitude is observed in both quantities. 
This is not surprising given the transition is first-order. There is a sudden shift of the energies and thus thermal occupation of the magnon-polaron bands,  in addition to the abrupt change of the topological structure.

In the antiparallel phase, as the energy of the spin-up band is lowered by the increasing field, it contributes more to the thermally driven transport properties dictated by the bosonic occupation. As the field increases above the critical field of the metamagnetic transition, only spin-down magnons are present, which split due to the the interlayer interaction instead. Across the transition, the dominating topological feature in thermal effects therefore switches from the $(1,-1)$ structure associated with spin-up to the $(-1,1)$ structure associated with spin-down, as discussed above. $\kappa_T$ is only sensitive to the band topology and thus flips sign, while $\alpha_{xy}^S$ also accounts for the spin and does not change sign. This argument works more directly for the semiclassical expression~\cite{cheng2016,park2020} of $\alpha_{xy}^S$, where the spin Berry curvature is replaced by Berry curvature multiplying spin weight. We have checked that the two give qualitatively consistent results, although the latter becomes compromised in the absence of spin conservation.


\begin{figure}
    \centering
    \includegraphics[width=0.48\textwidth]{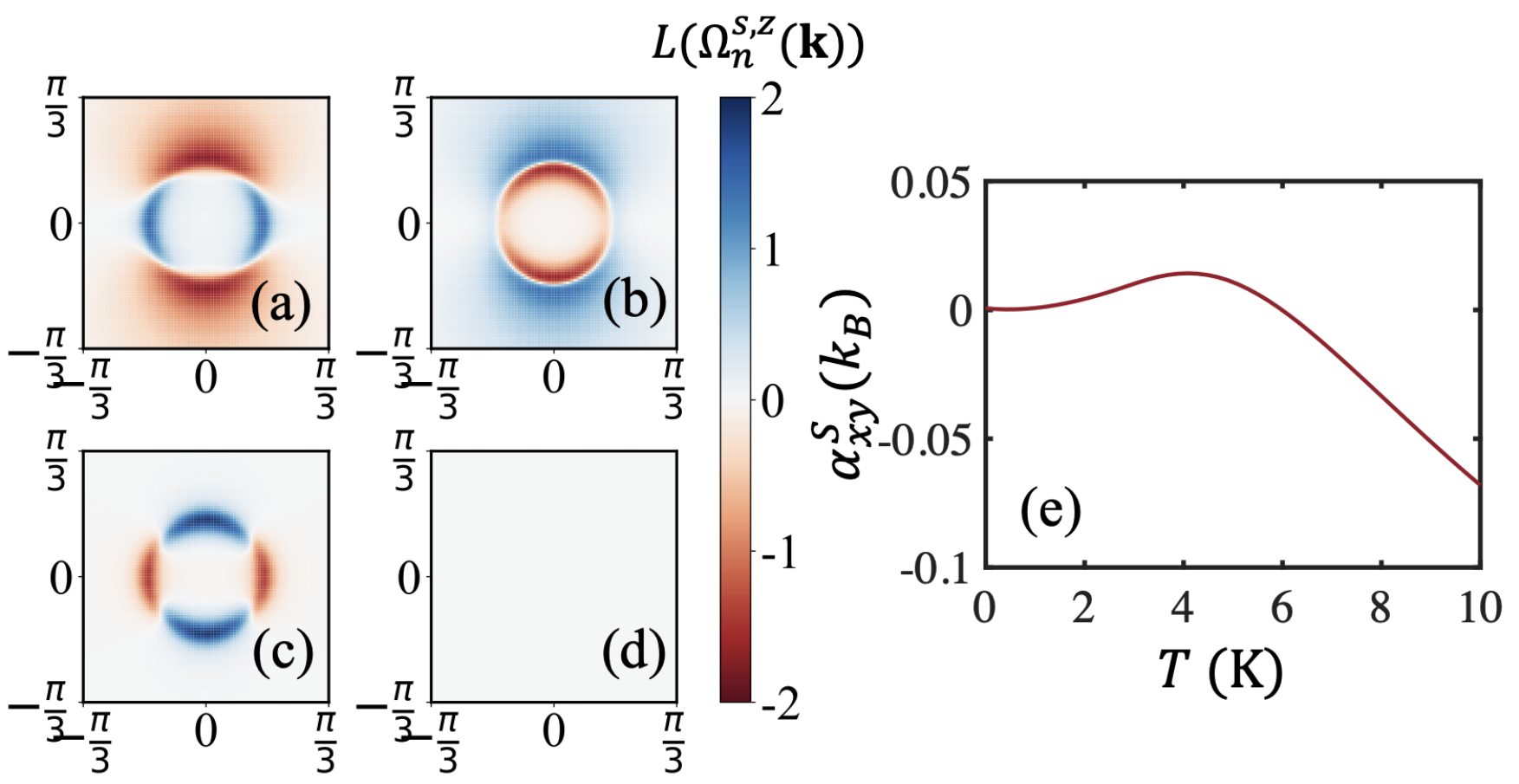}
    \caption{(a)-(d) The generalized spin Berry curvatures of band 1-4 and (e) the temperature dependence of the spin Nernst coefficient in the antiparallel phase under a small magnetic field $h=0.06\mathrm{meV}$, $L(\Omega_n^{s,z}(\vk))=\mathrm{sign}(\Omega_n^{s,z})\log (1+|\Omega_n^{s,z}|)$.}
    \label{fig:spin-bc}
    \vspace{-10pt}
\end{figure}

The spin conservation is completely invalidated when the applied magnetic field is weak, such that the energy splitting between the unperturbed magnon modes [Eq.~(\ref{eq:ap-magnon})] is smaller than the magnetoelastic coupling, i.e. the magnons are nearly degenerate everywhere in the Brillouin zone. The effective range of the hybridization involving both spin flavors is much extended~\cite{Go2022}. As shown in Figs.~\ref{fig:spin-bc}~(a-e), the spin Berry curvatures for the two lower bands remain finite in the area far away from the avoided crossings. Also, they exhibit sign changes in different regions within one band, breaking spatial rotational symmetry. The interplay between these factors and the thermal occupation lead to a smooth sign change of $\alpha_{xy}^S$ inside the antiparallel phase as a function of the magnetic field [Fig.~\ref{fig:transport}~(b)] and temperature [Fig.~\ref{fig:spin-bc}~(b)]. 
The spin Nernst effect does not require time-reversal breaking and can grow very large in the zero field limit according to Eq.~(\ref{eq:spin-bc}), due to the near degeneracy of band energies. In real materials, it is bounded by the scattering events that limit the lifetime of the magnon polarons.


We remark that our treatment is only valid at temperatures much below the N\'eel temperature. Temperature dependence of the transition field, scattering and many-body interacting effects~\cite{mook2021,li2023temperature}, disorder, magnetic domain wall, and other mesoscopic effects are not taken into account. The thermal Hall conductance resulted from our more detailed modeling of FeCl$_2$ is consistent with low-temperature experimental observations and theoretical calculations in Ref.~\cite{xu2023}, but does not resolve the discrepancies, consolidating the presence of other contributing mechanisms.
First-principle calculations show that atomically thin FeCl$_2$ films have a half-metallic behavior~\cite{feng2018robust,zhou2020atomically}. The interplay between topological magnon polarons and the itinerant electrons may play a role in the transverse thermal transport~\cite{watzman2016}. 
Our results also qualitatively apply to systems with more layers and other vdW magnets with antiferromagnetically coupled ferromagnetic layers, several examples available in the MnBi$_2$Te$_4$(Bi$_2$Te$_3$)$_{n}$ family~\cite{he2020,deng2020quantum,hu2020van,ovchinnikov2021intertwined} and transition-metal trihalides~\cite{wang2011electronic,mcguire2015coupling,mcguire2017crystal,huang2017layer,cai2019atomically}. It would be interesting to observe a distinction in thermal Hall transport for even and odd numbers of layers in the antiparallel phase. 

Magnon-phonon couplings can also arise from the general dependence of spin interactions on the bond length, which is modulated by phonon modes, and typically more effectively by displacements along the bond direction~\cite{takahashi2016,zhang2019,park2019,park2020,ma2022}. For a collinear magnetic ground state, couplings linear in magnon operators, which rely on breaking spin rotational symmetry, can be obtained by expanding the Dzyaloshinskii-Moriya or magnetic dipolar interactions. Such effects can thus be significant in systems with strong anisotropic spin interactions.  In our case, the magnetoelastic coupling~\cite{thingstad2019,go2019,szhang2020,Shen2020,Bazazzadeh2021Magnetoelastic,Go2022} is reminiscent more of a single-ion magnetic anisotropy. The out-of-plane orientation of the easy axis and the magnon coupling with out-of-plane displacements both comply with the lattice symmetry. Interestingly, a very strong easy-axis anisotropy, as is the case of transition metal dihalides FeI$_2$, can lead to bound excitations of magnon pairs~\cite{bai2021}, which form a dispersive band of quadrupolar nature below the magnon continuum. A hybridization between phonon and these magnetic bound states may also induce topological thermal transport effects in a similar fashion.
Our work motivates further study of magnon-polaron transport in vdW magnets.




See \textit{Supplemetal Material} for the detailed derivation of the BdG Hamiltonian and its full spectrum.

\begin{acknowledgments}
We thank G. A. Fiete, G. Go, M. Hirschberger, and S. K. Kim for useful discussions.
Z.-X. Lin thanks the Max Planck Institute for Physics of Complex Systems for support from the internship program.
\end{acknowledgments}

\bibliographystyle{apsrev4-1}
\bibliography{manuscript.bib}

\onecolumngrid
\clearpage
\setcounter{equation}{0}
\renewcommand{\theequation}{S\arabic{equation}}
\setcounter{figure}{0}
\renewcommand{\thefigure}{S\arabic{figure}}
\appendix

{\centering
    \large{\textbf{{Supplementary Material}}} 
\par}

\vspace{10pt}

\begin{figure*}[b]
    \centering
    \includegraphics[width=\textwidth]{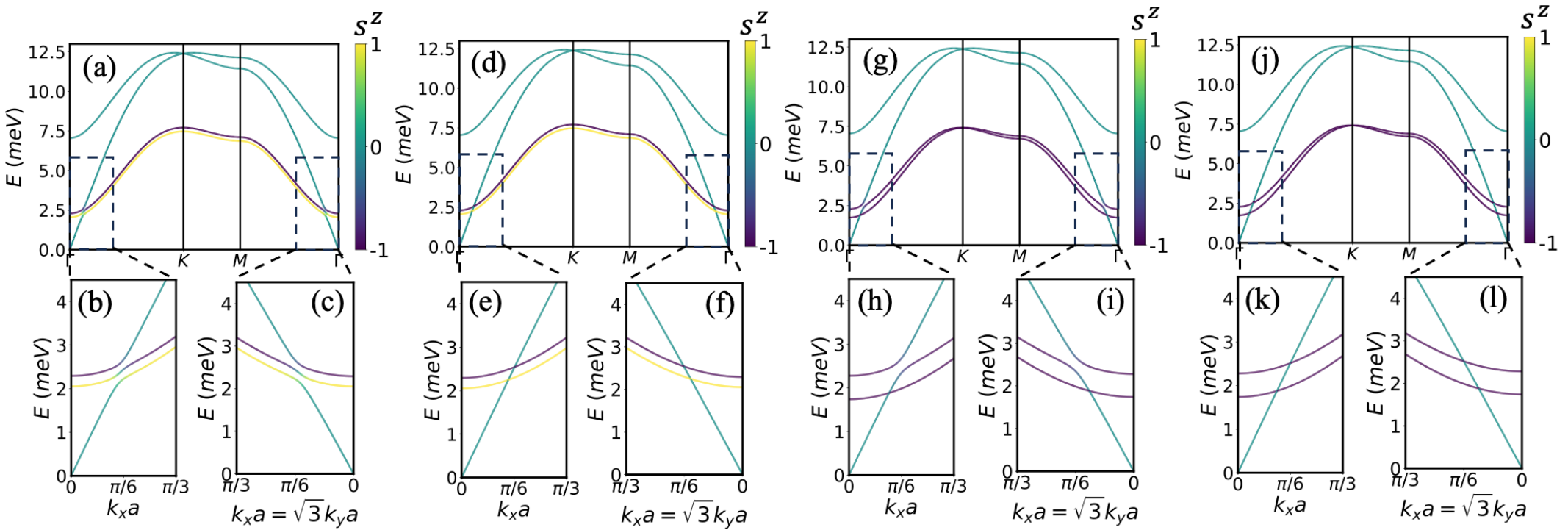}
    \caption{The band structure of the magnon-polaron excitations for (a)-(c) the antiparallel bilayer under the external field $h=0.12\ \mathrm{meV}$, in the presence of the magnetoelastic coupling strength $\kappa=0.18\ \mathrm{meV}/\AA$, (d)-(f) the antiparallel bilayer with $h=0.12\ \mathrm{meV}, \kappa=0$; (g)-(i) the parallel bilayer with $h=0.36\ \mathrm{meV}, \kappa=0.18\ \mathrm{meV}/\AA$, and (j)-(l) for the parallel bilayer with $h=0.36\ \mathrm{meV}, \kappa=0$. The color represents spin weight $s^z$ for each band at different momenta. 
    We adopted the parameters motivated by FeCl$_2$, $S=1$, $J=0.6$~meV, $K=2.0$~meV, $\hbar \omega_0 = 3.8$~meV, $J'=0.09$~meV,  and $\lambda'/\lambda = 0.58$. }
    \label{fig:bands}
\end{figure*}

In this supplementary material, we show the detailed derivation of the BdG Hamiltonian and the magnon-polaron band structure (Fig.~\ref{fig:bands}). We first express our Hamiltonian $H=(1/2)\sum_{\vk}\tilde{\phi}_{\vk}\tilde{\mathcal{H}}_{\vk}\tilde{\phi}_{\vk}$
in the basis $\tilde{\phi}_{\vk}=(a_{\vk},b_{\vk},a_{-\vk}^{\dagger},b_{-\vk}^{\dagger},c_{A,\vk},c_{B,\vk}, $ $c_{A,-\vk}^{\dagger},c_{B,-\vk}^{\dagger})^{T}$.
In this basis, the effective Hamiltonian $\tilde{\mathcal{H}}_{\vk}$
writes,
\begin{equation}
\tilde{\mathcal{H}}_{\vk}=\begin{pmatrix}\mathcal{H}_{\text{m}} & \mathcal{H_{\text{me}}}\\
\mathcal{H_{\text{me}}^{\dagger}} & \mathcal{H}_{\text{e}}
\end{pmatrix}.
\end{equation}

To derive the magnon Hamiltonian $\mathcal{H}_{m}$, we begin with the magnetic sector $H_\text{m}$ in the real space (Eq. \ref{eq:full-Hamiltonian}),
\begin{equation}
    H_\text{m} 
    \!=\! -J \!\!\!\!\! \sum_{ \langle \text{intra}\; ij  \rangle} \!\!\!\!\!
    \mathbf{S}_i \cdot \mathbf{S}_j
    + J' \!\!\!\!\! \sum_{ \langle  \text{inter}\; ij \rangle} \!\!\!\!\!
    \mathbf{S}_i \cdot \mathbf{S}_j
    - \frac{K}{2} \! \sum_i \! \left(S_i^z\right)^2
    - h \! \sum_i \! S_i^z.
\label{eq:magnetic-Hamiltonian}
\end{equation}
where $h \!=\! \mu_B g_\parallel B$ denotes the Zeeman energy with the Bohr magneton $\mu_B$ and the Land{\'e} g-factor $g_\parallel$ along the easy axis. By taking Holstein-Primakoff, we can obtain the magnon Hamiltonian in the momentum space. For the antiparallel bilayer, we have,
\begin{equation}
\mathcal{H}_{\text{m}}^{\text{AP}}=\begin{pmatrix}\epsilon_{+}^{m} & 0 & 0 & SJ'\gamma_{\vk}\\
0 & \epsilon_{-}^{m} & SJ'\gamma_{\vk}^{\ast} & 0\\
0 & SJ'\gamma_{\vk} & \epsilon_{+}^{m} & 0\\
SJ'\gamma_{\vk}^{\ast} & 0 & 0 & \epsilon_{-}^{m}
\end{pmatrix},
\end{equation}
where $\epsilon_{\pm}^{m}=S(JC_{\vk}+3J'+K)\pm h,C_{\vk}=6-2\cos k_{x}a-4\cos(k_{x}a/2)\cos(\sqrt{3}k_{y}a/2),\gamma_{\vk}=\exp(-i\sqrt{3}k_{y}a/3)+2\cos(k_{x}a/2)\exp(i\sqrt{3}k_{y}a/6)$, and the bosonic operator $a_{\vk}^\dagger$ $(b_{\vk}^\dagger)$ creates a magnon excitation carrying spin down (up) $s^z=-1 (+1)$  in layer A (B). And for the parallel bilayer, the magnon Hamiltonian takes the form of, 
\begin{equation}
\mathcal{H}_{\text{m}}^{\text{P}}=\begin{pmatrix}\epsilon_{+}^{m} & SJ'\gamma_{\vk} & 0 & 0\\
SJ'\gamma_{\vk}^{\ast} & \epsilon_{+}^{m} & 0 & 0\\
0 & 0 & \epsilon_{+}^{m} & SJ'\gamma_{\vk}\\
0 & 0 & SJ'\gamma_{\vk}^{\ast} & \epsilon_{+}^{m}
\end{pmatrix}.
\end{equation}
where both $a_{\vk}^\dagger$ and $b_{\vk}^\dagger$ create a spin-down magnon with $s^z=-1$. 

In the real space, the elastic part is written as, 
\begin{equation} 
    H_\text{e} \!=\! \frac{1}{2M} \! \sum_i \left(p^z_i\right)^2 
    + \frac{\lambda}{2} \!\!\!\! \sum_{ \; \langle \text{intra}\; ij \rangle} \!\!\!\!\! \left (u^z_i - u^z_{j}\right)^2
    +\frac{\lambda'}{2} \!\!\!\! \sum_{ \, \langle \text{inter}\; ij \rangle} \!\!\!\!\!  \left(u^z_i - u^z_{j}\right)^2,
\label{eq:phonon-Hamiltonian}
\end{equation}
where $u_{i}^z$ denotes the out-of-plane displacement on site $i$. Fourier transform yields
\begin{equation} 
    H_\text{e} = \frac{1}{2}\sum_{\vk} (u_{A,\vk},u_{B,\vk},p_{A,\vk},p_{B,\vk}) \tilde{\mathcal{H}}_\text{e} (u_{A,\vk},u_{B,\vk},p_{A,\vk},p_{B,\vk})^T, 
\end{equation}
with
\begin{equation}
\tilde{\mathcal{H}}_{\text{e}}=\begin{pmatrix}\lambda C_{\vk}+3\lambda' & -\lambda'\gamma_{\vk} & 0 & 0\\
-\lambda'\gamma_{\vk}^{\ast} & \lambda C_{\vk}+3\lambda' & 0 & 0\\
0 & 0 & 1/M & 0\\
0 & 0 & 0 & 1/M
\end{pmatrix}.
\end{equation}
To obtain the BdG Hamiltonian with bosonic commutation relation, we
introduce the following transform,
\begin{equation}
u_{i,\vk}=\sqrt{\hbar/2}(\lambda M)^{-1/4}(c_{i,\vk}+c_{i,-\vk}^{\dagger})\quad p_{i,\vk}=-i\sqrt{\hbar/2}(\lambda M)^{1/4}(c_{i,-\vk}-c_{i,\vk}^{\dagger})\quad(i=A,B).
\end{equation}
Expressing the elastic sector in the bosonic basis yields,
\begin{equation}
\mathcal{H}_{\text{e}}=\frac{\hbar}{2}\sqrt{\frac{\lambda}{M}}\begin{pmatrix}1+C_{\vk}+3\lambda'/\lambda & -\lambda'\gamma_{\vk}/\lambda & -1+C_{\vk}+3\lambda'/\lambda & -\lambda'\gamma_{\vk}/\lambda\\
-\lambda'\gamma_{\vk}^{\ast}/\lambda & 1+C_{\vk}+3\lambda'/\lambda & -\lambda'\gamma_{\vk}^{\ast}/\lambda & -1+C_{\vk}+3\lambda'/\lambda\\
-1+C_{\vk}+3\lambda'/\lambda & -\lambda'\gamma_{\vk}/\lambda & 1+C_{\vk}+3\lambda'/\lambda & -\lambda'\gamma_{\vk}/\lambda\\
-\lambda'\gamma_{\vk}^{\ast}/\lambda & -1+C_{\vk}+3\lambda'/\lambda & -\lambda'\gamma_{\vk}^{\ast}/\lambda & 1+C_{\vk}+3\lambda'/\lambda
\end{pmatrix}.
\end{equation}

Then we consider the symmetry-allowed magnetoelastic couplings~\cite{kittel1958,thingstad2019,go2019,szhang2020} on the triangular lattice, 
\begin{equation}\label{eq:me-Hamiltonian}
    H_\text{me} = \kappa \sum_{i} \sum_\delta S^z_i (\mathbf{S}_i \cdot \hat{\mathbf{e}}_\delta)(u_i^z-u_{i+\delta}^z),
\end{equation}
where $\kappa$ is the magnetoelastic coupling constant, $\delta$ iterates over the intralayer nearest neighbors of site $i$, and $\hat{\mathbf{e}}_\delta$ is the unit vector pointing along the bond. This term originates from spin-orbit coupling and breaks the conservation of $s^z$.

This magnetoelastic energy can also be expressed in $\tilde{\phi}_{\vk}$ basis,
\begin{equation}
\mathcal{H}_{\text{me}}^{\text{AP}}=\kappa\sqrt{S^{3}\hbar}(\lambda M)^{-1/4}\begin{pmatrix}-\nu_{1\vk}+\nu_{2\vk} & 0 & -\nu_{1\vk}+\nu_{2\vk} & 0\\
0 & \nu_{1\vk}+\nu_{2\vk} & 0 & \nu_{1\vk}+\nu_{2\vk}\\
-\nu_{1\vk}-\nu_{2\vk} & 0 & -\nu_{1\vk}-\nu_{2\vk} & 0\\
0 & \nu_{1\vk}-\nu_{2\vk} & 0 & \nu_{1\vk}-\nu_{2\vk}
\end{pmatrix},
\end{equation}
and,
\begin{equation}
\mathcal{H}_{\text{me}}^{\text{P}}=\kappa\sqrt{S^{3}\hbar}(\lambda M)^{-1/4}\begin{pmatrix}-\nu_{1\vk}+\nu_{2\vk} & 0 & -\nu_{1\vk}+\nu_{2\vk} & 0\\
0 & -\nu_{1\vk}+\nu_{2\vk} & 0 & -\nu_{1\vk}+\nu_{2\vk}\\
-\nu_{1\vk}-\nu_{2\vk} & 0 & -\nu_{1\vk}-\nu_{2\vk} & 0\\
0 & -\nu_{1\vk}-\nu_{2\vk} & 0 & -\nu_{1\vk}-\nu_{2\vk}
\end{pmatrix}.
\end{equation}
where $\nu_{1\vk} = i\sin k_x a +i \sin (k_x a/2) \cos (\sqrt{3}k_y a/2), \nu_{2\vk} = \sqrt{3} \cos (k_x a/2) \sin (\sqrt{3}k_y a/2)$. 

At the final step, we rearrange our original basis $\tilde{\phi}_{\vk}$
into the canonical BdG basis $\phi_{\vk}=(a_{\vk},b_{\vk},c_{A,\vk},c_{B,\vk},a_{-\vk}^{\dagger},b_{-\vk}^{\dagger},c_{A,-\vk}^{\dagger},c_{B,-\vk}^{\dagger})^{T}$.
To this end, we perform a global basis transformation, 
\begin{equation}
\mathcal{H}_{\vk}=P\tilde{\mathcal{H}}_{\vk}P^{\dagger},
\end{equation}
with
\[
P=\begin{pmatrix}\mathbf{1}_{2\times2} & \mathbf{0}_{2\times2} & \mathbf{0}_{2\times2} & \mathbf{0}_{2\times2}\\
\mathbf{0}_{2\times2} & \mathbf{0}_{2\times2} & \mathbf{1}_{2\times2} & \mathbf{0}_{2\times2}\\
\mathbf{0}_{2\times2} & \mathbf{1}_{2\times2} & \mathbf{0}_{2\times2} & \mathbf{0}_{2\times2}\\
\mathbf{0}_{2\times2} & \mathbf{0}_{2\times2} & \mathbf{0}_{2\times2} & \mathbf{1}_{2\times2}
\end{pmatrix}
.\]

\begin{figure*}[!h]
    \centering
    \includegraphics[width=0.65\textwidth]{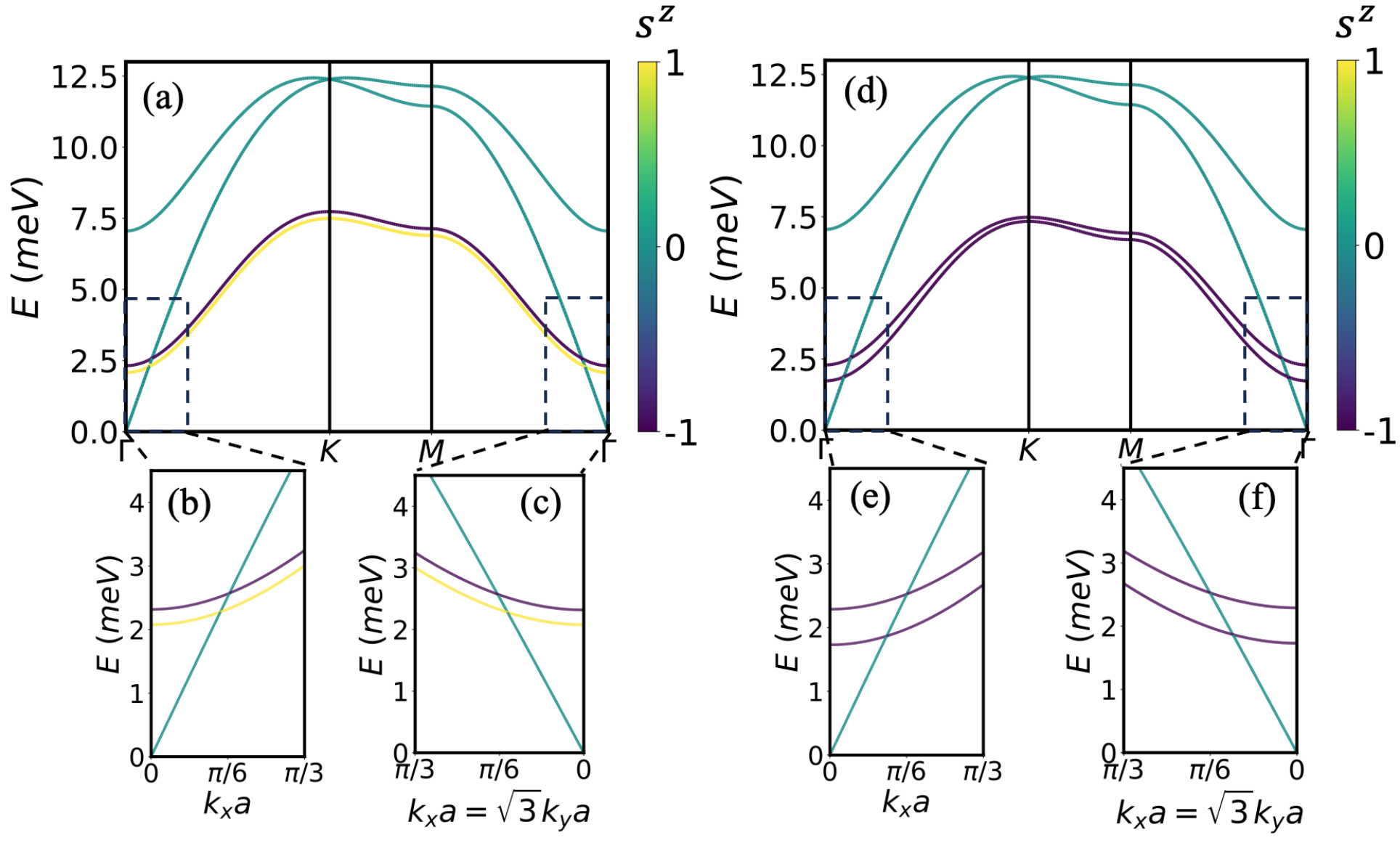}
    \caption{The band structure of the magnon-polaron excitations in absence of  the magnetoelastic coupling $\kappa=0$ with staggered field $\Delta h = 0.2h$ for (a)-(c) the antiparallel bilayer under the external field $h=0.12\ \mathrm{meV}$, and (d)-(f) the parallel bilayer with $h=0.36\ \mathrm{meV}$;  The color represents spin weight $s^z$ for each band. 
    Same parameters are used as in Fig. \ref{fig:bands}}
    \label{fig:bandsk=0}.
\end{figure*}

In the calculation of the band topology, we have introduced a small staggered magnetic filed $\Delta h$ between two layers. As shown in Fig. \ref{fig:bandsk=0}, in the non-interacting case ($\kappa=0$), the staggered field alone cannot induce avoided crossings between magnon and phonon branches.

\end{document}